# Electrical charge state manipulation of single silicon vacancies in a silicon carbide quantum optoelectronic device

**Authors:**

Matthias Widmann[1*], Matthias Niethammer[1], Dmitry Yu. Fedyanin[2], Igor A. Khramtsov[2], Torsten Rendler[1], Ian D. Booker[3], Jawad Ul Hassan[3], Naoya Morioka[1], Yu-Chen Chen[1], Ivan G. Ivanov[3], Nguyen Tien Son[3], Takeshi Ohshima[4], Michel Bockstedte[5,6], Adam Gali[7,8], Cristian Bonato[9], Sang-Yun Lee[1,10†] & Jörg Wrachtrup[1]

**Affiliations:**

[1] 3. Physikalisches Institut and Research Center SCOPE and Integrated Quantum Science and Technology (IQST), University of Stuttgart, Pfaffenwaldring 57, 70569 Stuttgart, Germany

[2] Laboratory of Nanooptics and Plasmonics, Moscow Institute of Physics and Technology, 9 Institutsky Lane, 141700 Dolgoprudny, Russian Federation

[3] Department of Physics, Chemistry and Biology, Linköping University, SE-58183 Linköping, Sweden

[4] National Institutes for Quantum and Radiological Science and Technology, Takasaki, Gunma 370-1292, Japan

[5] Department Chemistry and Physics of Materials, University of Salzburg, Jakob-Haringer-Str. 2a, 5020 Salzburg, Austria

[6] Solid State Theory, University of Erlangen-Nuremberg, Staudstr. 7B2, 91058 Erlangen, Germany

[7] Wigner Research Centre for Physics, Hungarian Academy of Sciences, PO. Box 49, H-1525 Budapest, Hungary

[8] Department of Atomic Physics, Budapest University of Technology and Economics, Budafoki út 8., H-1111 Budapest, Hungary

[9] Institute of Photonics and Quantum Sciences, SUPA, Heriot-Watt University, Edinburgh EH14

4AS, UK

[10] Center for Quantum Information, Korea Institute of Science and Technology, Seoul, 02792, Republic of Korea

*m.widmann@pi3.uni-stuttgart.de

†sangyun.lee236@gmail.com

Colour centres with long-lived spins are established platforms for quantum sensing and quantum information applications. Colour centres exist in different charge states, each of them with distinct optical and spin properties. Application to quantum technology requires the capability to access and stabilize charge states for each specific task. Here, we investigate charge state manipulation of individual silicon vacancies in silicon carbide, a system which has recently shown a unique combination of long spin coherence time and ultrastable spin-selective optical transitions. In particular, we demonstrate charge state switching through the bias applied to the colour centre in an integrated silicon carbide opto-electronic device. We show that the electronic environment defined by the doping profile and the distribution of other defects in the device plays a key role for charge state control. Our experimental results and numerical modelling evidence that control of these complex interactions can, under certain conditions, enhance the photon emission rate. These findings open the way for deterministic control over the charge state of spin-active colour centres for quantum technology and provide novel techniques for monitoring doping profiles and voltage sensing in microscopic devices.

Individual spins associated with quantum emitters in semiconductors are an established platform for quantum metrology and quantum information processing [1-5]. The possibility to manipulate individual spins builds on the capability to control the number of charges in a system, at the level of single electrons or single holes. This has been achieved with great success in the case of semiconductor quantum dots, through the Coulomb blockade effect [6-8]. Alternatively, colour centres can provide a system where individual spins can be controlled and detected, even at room temperature. Colour centres can exist in different charge states, each with a specific electronic structure featuring unique optical and spin properties. For example, the negative charge state of the nitrogen-vacancy (NV) centre in diamond hosts a coherent electronic spin which can be polarized and readout optically [1]. These properties have been exploited for quantum sensing [9,10,2,3] with nanoscale spatial resolution [11] and for seminal demonstrations of quantum networking [12–14]. Techniques have been developed to stabilize the colour centre charge state [15–18], as its fluctuations due to either noisy environment in solids or applied electromagnetic fields for control and readout is responsible for inefficiency in various applications [19–21,22]. Undesired switching to a different charge state precludes interfacing the electronic spin to photons. The fidelity of spin-photon interfacing can be preserved by triggering the experiment to start only when the colour centre is in the required charge state [14,23]. This, however, reduces protocol efficiency, decreasing the overall quantum communication rate [14]. In some applications, the possibility to switch between different charge states can enable novel functionalities, such as protecting a nuclear spin quantum memory by converting the colour centre to a spin-less charge state [24,25]. In general, precise control the charge state of the spin-active colour centre enables selecting the optimal properties relevant for the specific task [26–29].

Electrical control, by the bias applied through an electronic device, is a convenient and potentially deterministic way to access and manipulate any available charge state of a colour centre [29,30]. However, this is difficult in insulators and many wide-bandgap semiconductors like diamond. In this respect, silicon carbide (SiC) is a promising alternative since it uniquely combines the availability of several different colour centres featuring excellent quantum properties [30–35], with doping over a wide range of carrier densities [36,37], n-type as well as p-type. In addition, SiC features mature CMOS-compatible fabrication processes [38], which is a great benefit for scalable applications.

In this work, we focus on the single silicon vacancy ($V_{Si}$) in 4H-SiC and demonstrate electrical switching between the negatively-charged ($V_{Si}^{(-)}$) and the neutral ($V_{Si}^{(0)}$) charge states. The $V_{Si}^{(-)}$ has recently gained attention for its long spin coherence times [32,35,39–41], a strong optically-detected spin signal at cryogenic temperature [42,43,44], and ultrastable spin-selective optical transitions [44,45]. This combination of properties make it an extremely promising system to demonstrate efficient spin-to-photon interfacing for quantum networking [1]. However, very little is known about how the $V_{Si}$ charge state can be established and what the charge conversion mechanisms are. Here, we integrate single $V_{Si}$ centres into the intrinsic region of a 4H-SiC p-i-n diode and experimentally demonstrate electrical switching of the charge state of a single $V_{Si}$ by controlling the applied bias. In addition, enhancement of the photon emission is observed as well under specific applied bias values and optical excitation energies. To understand the microscopic nature of the phenomena observed experimentally, we present and test a model which reveals a complex interplay between the quasi-Fermi level tuning and optical excitation of the vacancy and other nearby defects.

**Charge state switching of the silicon vacancy**

The SiC p-i-n diode structure (see Fig. 1a,b) is grown by chemical vapour deposition (CVD) and consists of highly nitrogen-doped n-type and aluminum (Al)-doped p-type regions embedding a 50-μm-thick intrinsic layer. This intrinsic region is slightly p-type due to the residual Al and boron (B) impurities (see Methods).

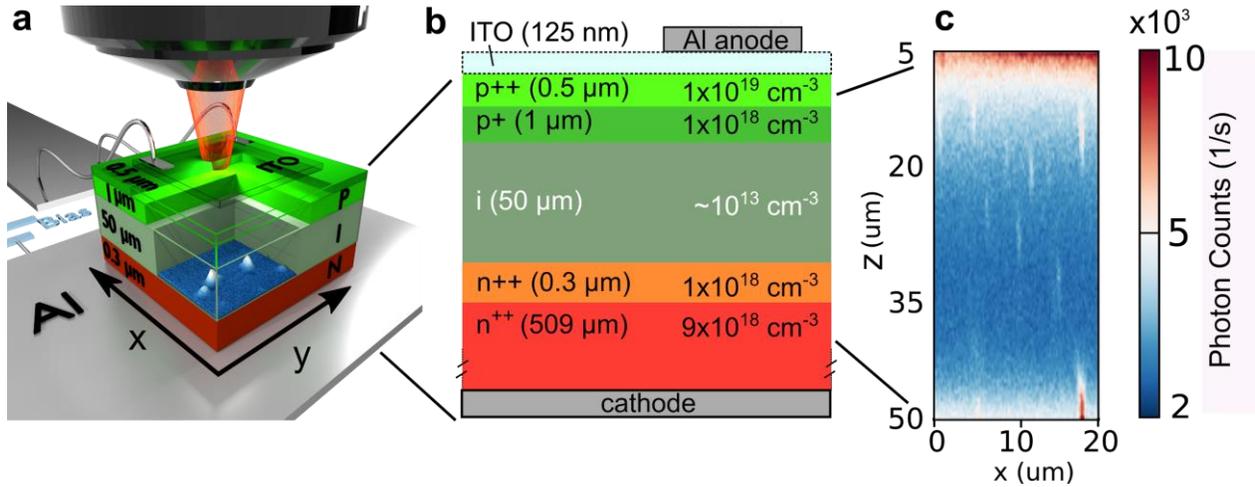

**Figure 1. $V_{Si}$ in SiC p-i-n junction device.** (a) Schematic of the p-i-n diode structure. The metallic base plate is Al. Red and green layers are the heavily doped n- and p-type layer, respectively. The pastel green layer is the intrinsic region, which is slightly p-doped. The ITO layer, which requires to form a transparent electrical contact, is on top of the p-type layer. (b) Detailed schematic of the device showing the thickness and the doping concentrations of each layer. (c) Room temperature confocal scan through the intrinsic layer.

In the first experiment, we perform a two-dimensional confocal scan of the device across the growth direction (z-axis) and one lateral axis (x-axis), at zero applied voltage (Fig. 1c). Using optical excitation at a wavelength of 730 nm (1.70 eV), we find isolated emitters across the intrinsic layer, which are identified as silicon vacancies in the negatively-charged state ($V_{Si}^{(-)}$) at the cubic lattice site (k) (see Methods). The optically detected spin Rabi oscillations of a single $V_{Si}^{(-)}$ as shown in Fig. 2d (see Supplementary Note 2) not only provide an evidence for $V_{Si}^{(-)}$ but also demonstrate that the capability of coherent spin manipulation and readout is maintained in the tested junction device. In the following we focus on the depletion region near the i-n junction of the diode structure. Strong band bending in this region gives the possibility to electrically control the charge states of the $V_{Si}$ centre simply when applying different bias voltages. We find that, while $V_{Si}^{(-)}$ in the intrinsic layer do not show significant changes in their density, the $V_{Si}$ centre near the i-n interface strongly respond to the applied bias: Fig. 2a shows confocal raster scans of the same x-y plane near the i-n interface (which is at a depth of about 47 µm) under reverse, zero and forward biases. At the reverse bias, several emitters are turned on, while the forward bias turns off the emitters that are bright at zero bias. We attribute this to switching of the charge state of the $V_{Si}$ from single negative to other dark charge states, which will be discussed in the subsequent sections.

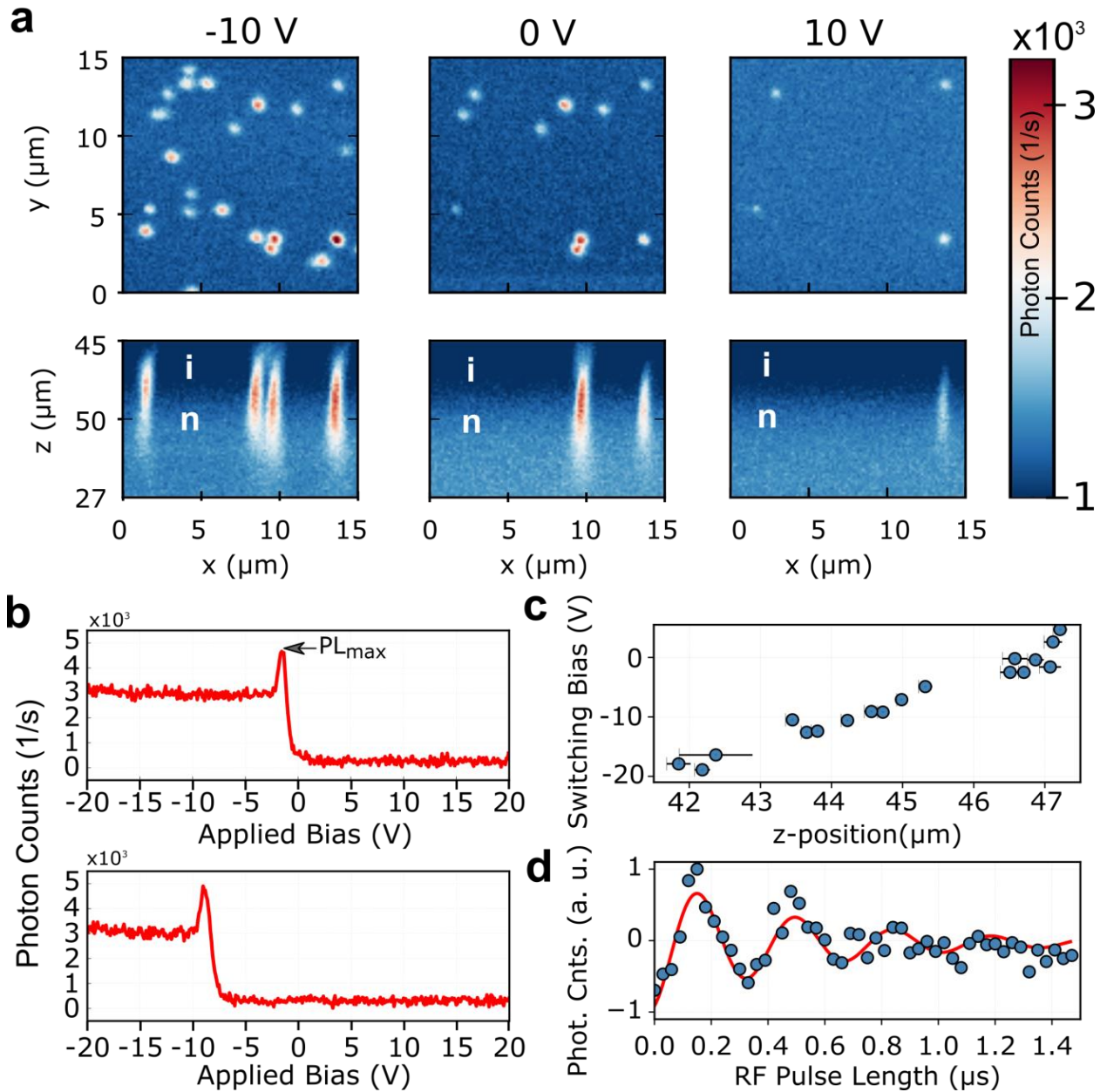

**Figure 2. Electrical charge-state switching of single silicon vacancies.** (a) Confocal scans at three bias voltages obtained at 1.5 mW (λ = 730 nm) optical excitation illustrating the charge state switching of $V_{Si}$ centres near the i-n interface. The labels "i" and "n" indicate the intrinsic and n-type layers, respectively. (b) PL intensities of two selected single $V_{Si}$ centres as a function of the bias voltage, illustrating the electrical switching of the charge states. (c) Dependence of the SB for individual $V_{Si}$ centers on their position along the z-axis. Each circle indicates the position of the tested $V_{Si(-)}$ and the bias voltage at which the switching occurs. For every center, the SB is extracted from the PL intensity vs applied bias curve similar to that shown in panel (b). (d) Optically detected spin-Rabi oscillations of a single $V_{Si}$ plotted at the expected resonant RF frequency. For (b) and (c), the optical excitation power is 5.5 mW (λ = 660 nm). See the text for details.

We also find that the newly switched-on emitters at reverse bias are located at a slightly further distance from the n-type layer. To test if the switching depends on the position of the $V_{Si}$ centre, we monitor the PL intensity of each bright emitter, while sweeping the bias voltage. Two selected results are shown in Fig. 2b. The PL intensity is completely turned off at forward bias, while it is bright at reverse bias. A sharp increase in the PL intensity is observed at the bias value inducing the switching, namely the switching bias (SB). Fig. 2b shows that the two tested emitters show different SB. To check if there exists a relation between the SB and the emitter position, we perform several confocal z-scans around each emitter and find the exact z-position (see Supplementary Note 4). By repeating this procedure on many emitters, we obtain a relation between the depth and the SB of each emitter as shown in Fig. 2c. This plot demonstrates that a stronger reverse bias is necessary to switch on $V_{Si}^{(-)}$ emitters located farther from the i-n interface.

To test how optical excitation is related with the observations in Fig. 1 and 2, we test 18 $V_{Si}$ centres and monitor their PL intensity as a function of the applied voltage.

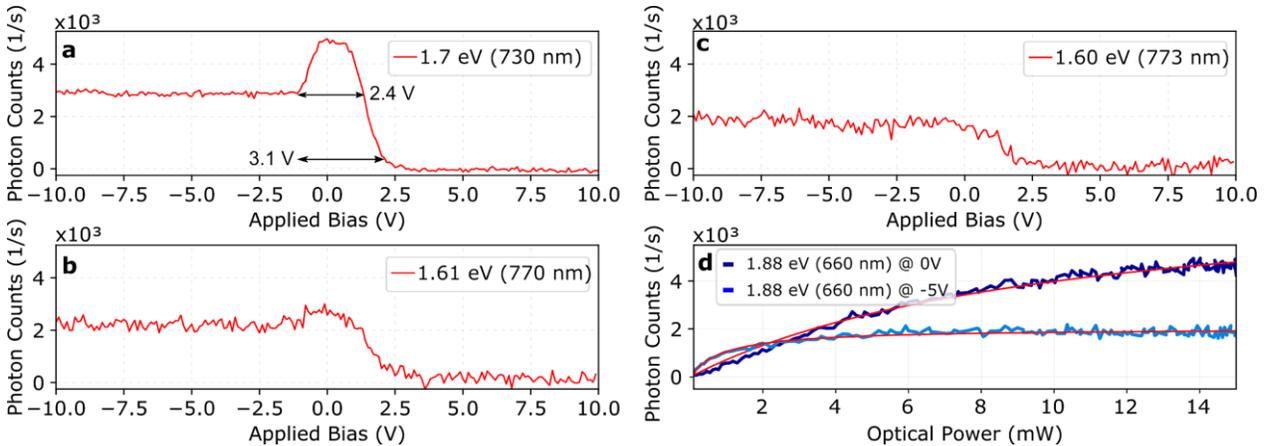

**Figure 3. Optical excitation dependence of charge state conversion.** (a-c) Bias dependent PL intensity curves under different excitation energies $\hbar\omega$: (a) 1.70 eV (730 nm), (b) 1.62 eV (765.5 nm), (c) 1.60 eV (773 nm). Arrows in a illustrate the width of increased PL (see text). (d) Optical saturation curves at 0 and -5 V, respectively, under 660 nm excitation. Solid lines are fit using $I = \alpha P/(\beta + P)$, where $\alpha$ and $\beta$ specify the saturated count rate and the saturation

power which are fit parameters. At 0 V, $\alpha$=5.2±0.3 kcts and $\beta$= 10±1 mW, and at -5V, $\alpha$= 2.0±0.3 kcts and $\beta$= 0.90±0.01 mW.

In Fig. 3a, the integrated PL intensity for the selected single $V_{Si}$ near the i-n interface is plotted versus the bias voltage for $\hbar\omega$ = 1.7 eV (λ = 730 nm). Again, charge-state switching is observed at around 0 V together with a sharp peak in the PL intensity. To understand the origin of the enhanced PL intensity around the SB (≈0 V for the $V_{Si}$ centre in Fig. 3), we vary the wavelength of the pump laser. Fig. 3b shows that similar curves are obtained at $\hbar\omega$ > 1.60 eV (λ < 773 nm). However, the peak in the PL intensity at the SB disappears at $\hbar\omega \lesssim$ 1.60 eV (λ ≳ 773 nm) (Fig. 3c) (see Supplementary Figure 6). Figure 3d shows that the PL intensity under optical saturation is 2 times stronger at 0 V than that at -5 V. Other tested $V_{Si}$ centre show the same behaviour, except that the SB is different for each center. These results suggest that an optical excitation energy larger than 1.60 eV triggers an additional process, resulting in an abrupt increase of the PL intensity, in addition to the electrical charge state switching. In the following, we discuss underlying mechanisms for the observations above.

**Charge states of the silicon vacancy**

In thermal equilibrium, the occupation of the neutral (0), single (-1), double (-2) and triple (-3) charged states are determined by the position of the Fermi level with respect to the valence band edge.

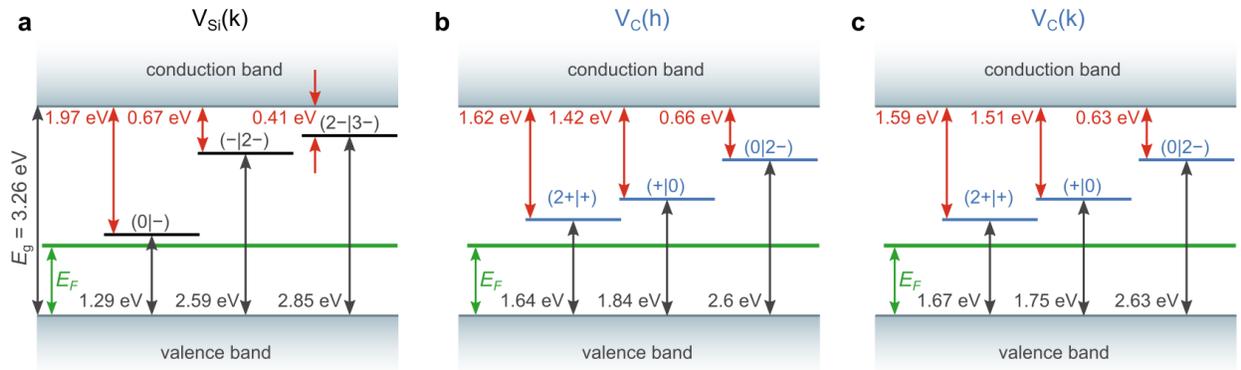

**Figure 4. Charge states of the $V_{Si}$ and $V_C$ centres.** Charge state transition levels of the (a) $V_{Si}$ at the cubic lattice site and (b,c) $V_C$ defects in 4H-SiC adopted from Refs. [26,46]. Note that we follow the recent assignment of the cubic and hexagonal defect of $V_{Si}$ in Ref. [47]. Generic values are provided in panels (b) and (c). The green line schematically shows the Fermi level. The black (red) arrows indicate the optical ionization energy towards higher (lower) charged states.

The known transition levels among the charge states of the $V_{Si}$, which have a deep-acceptor nature, are depicted in Fig. 4a. (see Methods and Supplementary Note 6). Figure 5a shows the simulated energy band diagram of the fabricated p-i-n diode in equilibrium (see Methods).

In the $p^{++}$- and $p^+$-type layers and in most of the intrinsic layer, which is slightly p-type, the Fermi level is below the (0|-) transition level. Hence, in these regions, the $V_{Si}$ is expected to be in the $V_{Si}^{(0)}$ state in equilibrium. However, under optical excitation, it may be ionized and converted into the $V_{Si}^{(-)}$ charge state. From the absorption spectrum of the $V_{Si}^{(0)}$ calculated using the CI-CRPA approach (see Supplementary Note 6), we obtain an optical excitation threshold 0.9 eV (1380 nm) for the conversion of the $V_{Si}^{(0)}$ into the $V_{Si}^{(-)}$. Because the optical excitation energy in this work is larger than $\hbar\omega$ = 1.58 eV (785 nm), the $V_{Si}^{(-)}$ can be observed.

In the region near the i-n junction, in equilibrium, the Fermi level crosses all three charge state transition levels of the $V_{Si}$ centre (see Fig. 5a). Hence, the $V_{Si}$ centre are expected to be in different charge states depending on their positions along the z-axis. For the charge state conversion in this region, we also have to take into account optical ionization. The conversion $V_{Si}^{(-)} \rightarrow V_{Si}^{(0)} + e^-$ is not likely since, it requires photon energy higher than 1.97 eV (629 nm), assuming only a single photon absorption process (see Fig. 4a), whereas the highest energy used in this study is 1.8 eV. In addition, optical ionization of other defects located in the vicinity of the studied $V_{Si}$ should also be considered, in particular, carbon vacancies ($V_C$), which are the most abundant intrinsic defects in 4H-SiC (Fig. 4b,c). Due to optical ionization, excess electrons and holes are released from defects to the conduction and valence bands. These carriers can be captured by $V_{Si}$ and other defects [48]. In our sample, the density of $V_C$ is in the range from $5\times10^{12}$ to $1\times10^{13}$ cm$^{-3}$ (see Methods). Since this density corresponds to an average distance between $V_C$ of ~300 nm, there is a good chance for a $V_C$ to be located near the studied $V_{Si}$. This complexity is further increased by the non-equilibrium induced by the applied bias as we will discuss in the next sections.

**Electrical control of the charge state**

To explain the mechanism of electrical switching between the charge states of the $V_{Si}$, we self-consistently simulate the 4H-SiC p-i-n diode shown in Fig. 1b (see Methods). Figure 5b shows the simulated energy band diagram at a bias voltage of -15 V applied to the device, which corresponds to a voltage drop across the p-i-n diode of -3 V due to non-ideality of the fabricated device (see Methods).

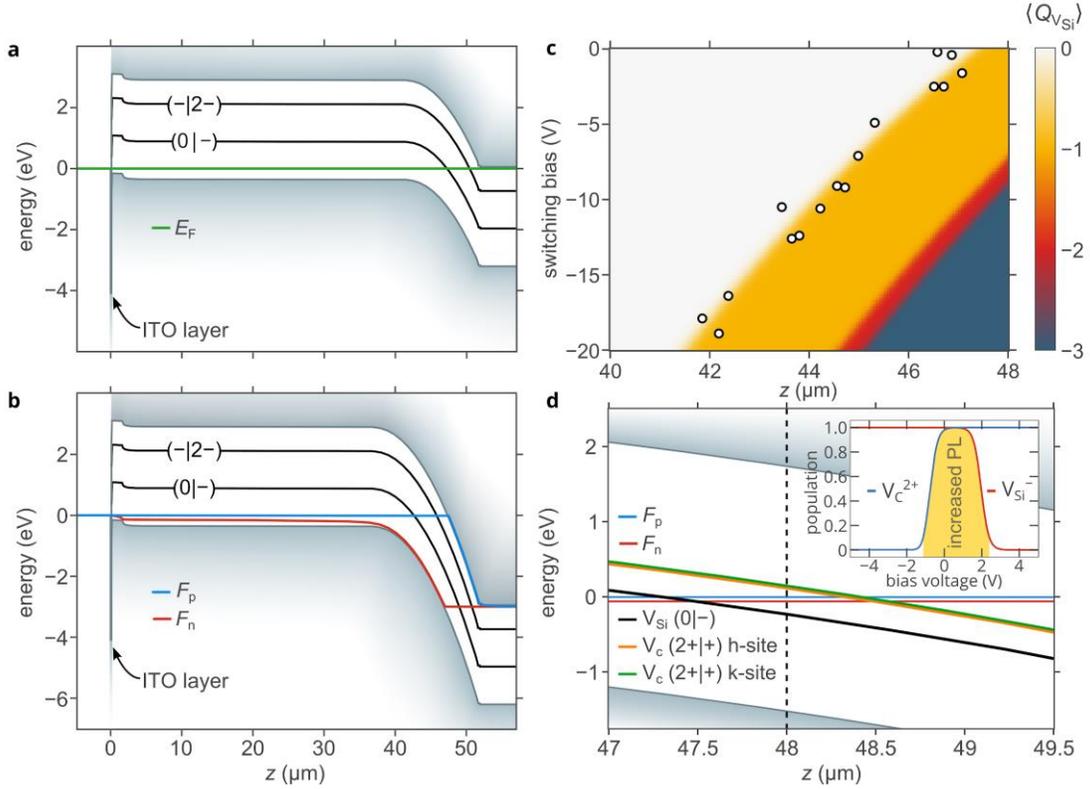

**Figure 5. Charge state conversion by applied bias.** (a,b) Simulated energy band diagrams of the fabricated device in thermal equilibrium (a) and at a reverse bias voltage of -15 V (b). Black solid lines show the positions of the charge transition levels (0|-) and (-|2-). The (2-|3-) transition level is not shown since it is located very close to the (-|2-) transition level. (c) Simulated evolution of the spatial distribution of the time-averaged charge (i.e., averaged occupation of the neutral, (-1), (-2) and (-3) charge states as denoted by the color scale) of single silicon vacancies <$Q_{VSi}$> with the reverse voltage applied to the device. Open dots represent the results of the experimental measurements shown in Fig. 2c. (d) Energy band diagram at around $z$ = 48 μm at $V$ = -0.3 V. Inset: Populations of the (-1) charge state of the silicon vacancy and (+2) charge state of the carbon vacancy at $z$ = 48.3 μm versus bias voltage. The yellow area shows the voltage range of the increased PL.

As the reverse bias increases, the band bending in the depletion region near the i-n junction increases and the depletion region expands towards the p-i junction. At the same time, the applied bias perturbs the carrier equilibrium , which splits the Fermi level $E_F$ into two quasi-Fermi levels $F_n$ and $F_p$ for electrons and holes, respectively. Accordingly, the occupation of the charge states cannot be identified as easily as in equilibrium. One has to closely consider the processes of electron and hole capture and release by the V$_{Si}$ in a manner similar to that used for the description of electroluminescence of color centres [34,49–51]. In the region where we observe switching of the charge state of the V$_{Si}$ (see Fig. 2c), we find that the density of holes is many orders of magnitude higher than that of electrons, even at significantly high voltages (see Supplementary Figure 9). In addition, there are no minority carriers in the reverse biased diode. Therefore, in the band bending region the occupation of the charge states of the $V_{Si}$ is mainly determined by the hole capture and hole release processes. This situation is the same as in a p-type material in equilibrium. Thus, we can use the same expressions for the charge states populations of the $V_{Si}$ as in equilibrium by replacing $E_F$ with $F_p$. In other words, the transition between the charge states occurs when the quasi-Fermi level for holes crosses the corresponding transition level. Figure 5c shows the corresponding calculations based on this quasi-equilibrium approach, predicting electrical switching between the $V_{Si}^{(0)}$ and $V_{Si}^{(-)}$ states. According to calculations (see Supplementary Note 6), $V_{Si}^{(0)}$ is expected to emit in the near-infrared, with photon energies less than the ionization threshold (~0.9 eV). Emission, however, is expected to be weak due to competing non-radiative processes. Since the detectors used in the experiment are not sensitive for wavelengths larger than ~1000 nm, switching to the $V_{Si}^{(0)}$ charge state corresponds to a suppression of the PL signal. Together with these simulation results, the data in Fig. 2b and Fig. 3 show that the p-i-n diode structures allows to switch the charge state between $V_{Si}^{(0)}$ and $V_{Si}^{(-)}$ by applying a moderate voltage.

**Reinitialization of the silicon vacancy**

Finally, we turn to the origin of the increased PL at around the switching bias as shown in Fig. 2b and Fig. 3. For simplicity, we focus on the particular $V_{Si}$ centre located at a depth of about 48 μm, for which the measurements are shown in Fig. 3. However, this analysis can be applied with no change to any observed $V_{Si}$ in the studied 4H-SiC diode. As discussed in the previous section, around the switching bias voltage the occupation of the $V_{Si}^{(0)}$ steadily decreases from 100% to 0 and the occupation of the $V_{Si}^{(-)}$ increases from 0 to 100% as the bias voltage decreases (see Fig. 3c). However, at optical excitation energies above 1.6 eV, we observe an increased PL rate in the voltage range from -1.1 V to 2.0 V (Fig. 3a,b). If the PL at voltages below -1.1 V corresponds to 100% occupation of the (-1) charge state, why does the PL rate at voltages between -1.1 V and 1.3 V become higher (see Fig. 3a,b)? This is counterintuitive, since the occupation of the negative charge state (-1) cannot increase further. To resolve this contradiction and find an additional process that can trigger an increase in the PL intensity, we emphasize that the increased PL of the $V_{Si}$ is accompanied with the charge state switching from (0) to (-1), and therefore we can expect that complex electron and hole capture and release processes may happen at around the switching bias. Additionally, the peak at the switching bias is observed only at optical excitation energies higher than 1.6 eV, which approximately coincides with the optical ionization threshold of $V_C^{(2+)} \rightarrow V_C^{(+)}+h$ (Fig. 4b,c). This suggests that, holes released by $V_C$ under optical excitation may be captured by the $V_{Si}^{(-)}$.

To understand how $V_C$ can affect the $V_{Si}$ charge state, we simulate the energy band diagram in the vicinity of the $V_{Si}$ at $V$ = -0.3 V applied to the device (see Methods), which corresponds to the voltage drop across the p-i-n diode of -0.06 V . Since at $z$ = 48 μm, $F_p$ lies above the $V_{Si}$ (0|-) level, the considered $V_{Si}$ is in the (-1) charge state. At the same time, at $z$=48 μm, $F_p$ is below the $V_C$ (2+|+) level. Accordingly, the $V_C$ at $z$ = 48 μm is in the (+2) charge state, so are all $V_C$ defects at $z$ < 48 μm. We assume a $V_C$ in the vicinity of the $V_{Si}$ can be ionized by the excitation laser, if $\hbar\omega \gtrsim 1.64$ eV (see Fig. 4b,c), and release a hole to the valence band, i.e., $V_C^{(2+)} \rightarrow V_C^{(+)}$+h. This hole can be captured by the considered $V_{Si}$, thus $V_{Si}^{(-)}$+h$\rightarrow V_{Si}^{(0)}$. As an increased PL is evident, the $V_{Si}^{(0)}$ rapidly returns back into $V_{Si}^{(-)}$. There are two possibilities for this transition. (I) The excitation laser can ionize the $V_{Si}^{(0)}$, which is possible at photon energies 1.6-1.8 eV as shown in Supplementary figure 7), and bring the $V_{Si}$ back to the $V_{Si}^{(-)}$ ground state. However, this cannot result in increased luminescence. (II) The $V_{Si}^{(0)}$ can capture an electron from the conduction band. This electron can be provided due to the non-ideality of the device, and consequently non-zero electron current, or by the photoionization of another defect located nearby the $V_{Si}$. The free electron is captured by the $V_{Si}$ into the $V_{Si}^{(-)}$ excited state [34,49], which then relaxes to the $V_{Si}^{(-)}$ ground state via photon emission increasing the luminescence rate dramatically [34,49–51], thus, $V_{Si}^{(-)}$ is re-initialized. This mechanism explains the experimentally observed threshold optical energy of 1.6 eV and supports that the release of a hole by $V_C$ ionization followed by its capture at the $V_{Si}^{(-)}$, which promotes the reinitialization of $V_{Si}^{(-)}$, is very likely. For a larger reverse bias, the band bending is steeper (see Fig. 5b), so is the $V_C$ (2+|+) level, which repeats the profile of the conduction and valence bands. Therefore, the point of intersection between the $V_C$ (2+|+) level and $F_p$ shifts toward the p-i interface. At $V$ < -1.4 V, all carbon vacancies at $z \geq$ 48 μm are in the (+1) charge state and consequently $V_C^{(2+)} \rightarrow V_C^{(+)}$+h is not possible. This essentially stops the reinitialization of the $V_{Si}^{(-)}$. Thus, according to our theoretical model and numerical simulations, for the $V_{Si}$ center at $z$ = 48 μm, the enhanced PL should be observed in the voltage range from -1.4 to 1.5 V, which is in good agreement with the experimental values (from -1.1 to 2.0 V). Even better coincidence with the experimental results can be obtained assuming the $V_{Si}$ to be at $z$ = 48.3 μm (see inset in Fig. 5d). In this case, the voltage range of the enhanced PL is from -1.1 to 2.4 V. Although the experimental observations can be qualitatively explained by our model, the proposed mechanism may not be the only explanation. Further understanding could be achieved by combining the method used in this study with other junction spectroscopic methods, such as deep level transient spectroscopy [52].

**Conclusions**

The results presented in this work show for the first time electrical manipulation of the charge states of single silicon vacancies centres in silicon carbide optoelectronic device. We demonstrate switching of the silicon vacancy in the i-region of the 4H-SiC p-i-n diode between the neutral and single negatively charged states. We also find that the optical ionization of the silicon vacancy and other nearby defects such as carbon vacancies play an important role in charge state switching. When the ionization of carbon vacancies re-initializes the silicon vacancy, we observe even an enhancement of the PL intensity of the silicon vacancy. Our work demonstrates not only a convenient way to control the charge state of atomic-scale defects in semiconductor quantum optoelectronic devices but also potential applications using the atomic-scale colour centres as a probe for local Fermi levels. This may open a new pathway to design efficient and robust quantum interfaces for quantum-repeater applications [44] and may improve efficiency of quantum internet protocols [14] since idling time in an unfavoured dark state can be minimized. The demonstrated method may be extended to other colour centres in silicon carbide [1] and similar materials [53,54], and used as an atomic-scale probe to characterize the distributions of defects states at the junctions of optoelectronic devices. By further optimization of the device structure and doping methods, much steeper changes of the PL intensities at around the switching bias will lead to new methods to sense electrostatic potentials at the nanoscale.

**Methods**

**Experimental setup**. All measurements were performed in a typical home-built confocal microscopy setup at room temperature. Laser diodes for fixed wavelength optical excitation and Ti:Sapphire laser as a tunable light source were used. A source and measurement (SMU) unit (Keithley 4200SCS) was used for applying a bias to the p-i-n structure. Note that the origin for the z-axis is set to the top surface of the sample. The depth in Fig. 1c is corrected to account for aberrations induced by refraction from the oil-SiC interface (see Supplementary Note 4)[55]. Because the spatial resolution of a confocal microscope is worse in the axial direction, fluorescent spots appear stretched along the *z*-axis in Fig. 1c and in Fig. 2a. In Fig. 1, 2, and 3, the integrated PL intensity in the spectral range 900-1050 nm is plotted. Full description for the experimental method can be found in the Supplementary Note 1.

**Sample preparation.** The SiC p-i-n diode structure used is grown by chemical vapour deposition (CVD) on a 4-degree off-cut 4H-SiC substrate. The 509-µm-thick n-type substrate is heavily doped with nitrogen (N) at a density of about $9\times10^{18}$ cm$^{-3}$. The p-i-n diode structure (see Fig. 1a,b) consists of highly N-doped n-type (~$9\times10^{18}$ cm$^{-3}$) and aluminium (Al)-doped p-type ($10^{18}$ cm$^{-3}$) regions embedding a 50-µm-thick intrinsic layer. This intrinsic region is slightly p-type due to the residual Al and boron (B) impurities (~low $10^{13}$ cm$^{-3}$). The fabricated SiC structure is placed on an Al base plate. The contact between the n-type substrate and the Al plate is provided by silver paste. For obtaining a p-type contact, a $p^{++}$-type layer is grown on top of the p-type layer. Next, a 200-nm-thick indium tin oxide (ITO) transparent conductive layer is deposited on top of the $p^{++}$-type layer (see Fig. 1b). Finally, Al electrodes are deposited on a small part of the ITO layer and wire-bonded to a large electrode on the sample holder by Al wires. The selected approach provides not only an electrical contact to the $p^{++}$-type layer but also allows optical access for excitation and monitoring of silicon vacancies deep in the intrinsic layer of the diode structure. Isolated silicon vacancies in the intrinsic layer are created by irradiating the fabricated structure with 2-MeV electron beams at a fluence of $5\times10^{12}$ cm$^{-2}$ before the deposition of the ITO layer. Because the silicon vacancies are created by high-energy electron irradiation (penetration depth ~ 2 mm), one can expect a homogeneous distribution of silicon vacancies over the whole intrinsic layer. See the Supplementary Note 3 for the density of the created silicon vacancies as a function of the electron irradiation dose. The density of $V_C$ defects is in the range from $5\times10^{12}$ to $1\times10^{13}$ cm$^{-3}$, which is measured by deep-level transient spectroscopy

**Identification of single silicon vacancies.** In 4H-SiC, the silicon vacancy can be created at the two inequivalent lattice sites, namely the hexagonal and cubic sites, called V1 and V2 centre, respectively [47]. Both defects possess very similar optical spectra with slightly different PL peaks. In this work, all the found silicon vacancies are the cubic silicon vacancy. The V2 centre can be identified by their characteristic photoluminescence (PL) spectrum in the spectral range 900-1050 nm [56] (data not shown) and optically detected magnetic resonance (ODMR) signature (see Fig. 2d). The studied emitters are verified to be single-photon using the $g^{(2)}$-function measurements in the Hanbury-Brown and Twiss configuration (data not shown). Although it is known that the negatively charged $V_{Si}$ can exist in n-type SiC such as in semi-insulating SiC [57], we have not found $V_{Si}^{(-)}$ centres in the n-type layers of the fabricated structure. Our studies have also not revealed the presence of the $V_{Si}^{(-)}$ in the $p^{++}$- and $p^{+}$-type layers ([Al]~$10^{19}$ cm$^{-3}$), which agrees well with the previous studies [57]. However, we clearly observed the $V_{Si}^{(-)}$ spectrum in the slightly p-type doped i-layer, which may be due to the charge state conversion via optical ionization, as discussed in the text.

**Charge state transition levels of the silicon vacancy.** We study the charge states of the $V_{Si}$ centre using hybrid density functional theory (hybrid-DFT) and the configuration interaction method CI-CRPA [58] (see Supplementary Note 6), which reveals the deep-acceptor nature of the silicon vacancy. The known transition levels among the charge states are shown in Fig. 4a. In thermal equilibrium, the occupation of the (0), (-1), (-2) and (-3) charged states are determined by the position of the Fermi level with respect to the valence band edge and can be found using the Gibbs distribution [59,60]. Since, the thermal energy $kT$ is much lower than the bandgap energy $E_g$=3.26 eV under ambient conditions, we can define the positions of the Fermi level at which the transition between the $V_{Si}^{(0)}$ and $V_{Si}^{(-)}$ charge states occurs: the Fermi level ($E_F$) in 4H-SiC should be 1.29 eV above the valence band edge [59]. In Fig. 4a, this transition level is labeled as (0|-). For $E_F$ > 2.59 eV, the defect turns into the $V_{Si}^{(2-)}$, and further into the $V_{Si}^{(3-)}$ for $E_F$ > 2.85 eV [59]. These levels are labeled as (-|2-) and (2-|3-), respectively. Since only the $V_{Si}^{(-)}$ centre is known to exhibit PL in the spectral region 900-1050 nm, silicon vacancies in the negative charge state can be detected as bright spots in the experiment.

**Numerical simulations of the silicon carbide diode.** The 4H-SiC p-i-n diode was numerically simulated using the self-consistent steady-state model which comprises the Poisson equation for the electric field and carrier densities, the semiconductor drift-diffusion equations and the carrier continuity equations for electrons and holes. These equations were supplemented by the appropriate boundary conditions (the top and bottom contacts were modeled as ideal ohmic contacts) and solved using the nextnano++ software. Since the density of acceptors $N_{Ai}$ and the acceptor compensation ratio $\eta_A$ by donor-type defects are the only unknown diode parameters, we assume $\eta_A$ to be 10% and slightly vary $N_{Ai}$ to better fit the experimental data. These and other parameters used in the simulations can be found in Supplementary Table 1. In addition, we take into account the non-ideality of the fabricated device. The measured current-voltage characteristic is linear at very high forward and negative voltages. The slopes at these voltages correspond to resistances of about 8 and 10 kΩ, respectively (see Supplementary Note 7). This non-ideality can be represented by an equivalent circuit with two resistors of 2 and 8 kΩ, which are connected to the p-i-n diode in parallel and series, respectively. The parallel resistance models possible parasitic conduction through the edges of the sample, while the series resistance accounts for non-ideal contacts and lead wires.

Note that the prediction of switching among $V_{Si}^{(-)}$, $V_{Si}^{(2-)}$ and $V_{Si}^{(3-)}$ may be more complex since it should occur at electron and hole densities of less than $10^{-20}$ cm$^{-3}$ (see Fig. 5c and Supplementary Figure 9). In these regimes, the densities of excess electrons and holes produced by optical ionization of $V_C$ and other defects are much higher. Therefore, in the i-layer, electrical switching among the $V_{Si}^{(-)}$, $V_{Si}^{(2-)}$ and $V_{Si}^{(3-)}$ states is more difficult.

**Acknowledgement**

The authors thank Brian Gerardot, Jaekwang Lee, Tsunenobu Kimoto, Helmut Fedder and Stefan Lasse for helpful discussions. This work is supported by the Ministry of Education and Science of the Russian Federation (8.9898.2017/6.7), the Russian Foundation for Basic

Research (19-57-12008), grant of the President of the Russian Federation (MK-2602.2017.9), Baden-Württemberg Stiftung Programm: Internationale Spitzenforschung, the ERC SMel and the BMBF BRAINQSENS, the Korea Institute of Science and Technology institutional program (2E29580, 2E27110), the Swedish Research Council (VR 2016-04068 and VR 2016-05362), the Carl Tryggers Stiftelse för Vetenskaplig Forskning (CTS 15:339), the Swedish Energy Agency (43611-1), the Knut and Alice Wallenberg Foundation (KAW 2018.0071), the National Quantum Technology Program (Grant No. 2017-1.2.1-NKP-2017-00001), National Excellence Program (Grant No. KKP129866), and EU QuantERA Nanospin project (Grant No. 127902) from the National Office of Research, Development and Innovation in Hungary, and the Engineering and Physical Sciences Research Council (EP/P019803/1 and EP/S000550/1) and the Networked Quantum Information Technologies Hub (Oxford) and JSPS KAKENHI (17H01056 and 18H03770). Supercomputer time was granted on the HPC cluster of the RRZE of the Friedrich-Alexander Universität, Erlangen-Nürnberg, the Doppler-Cluster of the Paris-Lodron University Salzburg.

**Author Contributions**

S-Y.L. conceived the conceptual idea; M.W. and S-Y.L. designed the charge state control experiment; S-Y.L. and N.T.S. designed the sample structure; M.W. performed the charge state control experiments; M.N., T.R., and S-Y.L. provided experimental assistance; J.U.H., I.G.I., and N.T.S. prepared the sample and characterized basic properties; T.O. performed electron irradiation; M.W., M.N., D.Y.F., I.A.K., N.M., C.B., and S-Y.L. analyzed the data; D.Y.F., and I.A.K. developed the theory and performed computations for charge state conversion; M.B. calculated the optical transitions of the studied defects; A.G. provided theoretical support; M.W., D.Y.F., M.B., C.B., N.M., T.R. and S-Y.L. wrote the manuscript; All authors discussed and commented on the manuscript.

**References**

1. Atatüre, M., Englund, D., Vamivakas, N., Lee, S.-Y. & Wrachtrup, J. Material platforms for spin-based photonic quantum technologies. *Nat. Rev. Mater.* **3**, 38–51 (2018).
2. Casola, F., van der Sar, T. & Yacoby, A. Probing condensed matter physics with magnetometry based on nitrogen-vacancy centres in diamond. *Nat. Rev. Mater.* **3**, 17088

(2018).

3. Degen, C. L., Reinhard, F. & Cappellaro, P. Quantum sensing. *Rev. Mod. Phys.* **89**, 035002 (2017).

4. Awschalom, D. D., Hanson, R., Wrachtrup, J. & Zhou, B. B. Quantum technologies with optically interfaced solid-state spins. *Nature Photonics* **12**, 516–527 (2018).

5. Zwanenburg, F. A. *et al.* Silicon quantum electronics. *Reviews of Modern Physics* **85**, 961–1019 (2013).

6. Hanson, R., Kouwenhoven, L. P., Petta, J. R., Tarucha, S. & Vandersypen, L. M. K. Spins in few-electron quantum dots. *Reviews of Modern Physics* **79**, 1217–1265 (2007).

7. Warburton, R. J. Single spins in self-assembled quantum dots. *Nature Materials* **12**, 483–493 (2013).

8. Brotons-Gisbert, M. *et al.* Coulomb blockade in an atomically thin quantum dot coupled to a tunable Fermi reservoir. *Nat. Nanotechnol.* **14**, 442 (2019).

9. Balasubramanian, G. *et al.* Nanoscale imaging magnetometry with diamond spins under ambient conditions. *Nature* **455**, 648–651 (2008).

10. Häberle, T., Schmid-Lorch, D., Karrai, K., Reinhard, F. & Wrachtrup, J. High-dynamic-range imaging of nanoscale magnetic fields using optimal control of a single qubit. *Phys. Rev. Lett.* **111**, 170801 (2013).

11. Hong, S. *et al.* Nanoscale magnetometry with NV centers in diamond. *MRS Bull.* **38**, 155–161 (2013).

12. Hensen, B. *et al.* Loophole-free Bell inequality violation using electron spins separated by 1.3 kilometres. *Nature* **526**, 682–686 (2015).

13. Kalb, N. *et al.* Entanglement distillation between solid-state quantum network nodes. *Science* **356**, 928–932 (2017).

14. Humphreys, P. C. *et al.* Deterministic delivery of remote entanglement on a quantum network. *Nature* **558**, 268–273 (2018).

15. Doi, Y. *et al.* Deterministic Electrical Charge-State Initialization of Single Nitrogen-Vacancy Center in Diamond. *Physical Review X* **4**, 011057 (2014).

16. Petráková, V. *et al.* Luminescence properties of engineered nitrogen vacancy centers in a close surface proximity. *physica status solidi (a)* **208**, 2051–2056 (2011).

17. Fávaro de Oliveira, F. *et al.* Tailoring spin defects in diamond by lattice charging. *Nat. Commun.* **8**, 15409 (2017).

18. Grotz, B. *et al.* Charge state manipulation of qubits in diamond. *Nat. Commun.* **3**, 729 (2012).

19. Bradac, C. *et al.* Observation and control of blinking nitrogen-vacancy centres in discrete nanodiamonds. *Nat. Nanotechnol.* **5**, 345–349 (2010).

20. Han, K. Y. *et al.* Dark state photophysics of nitrogen–vacancy centres in diamond. *New Journal of Physics* **14**, 123002 (2012).

21. Aslam, N., Waldherr, G., Neumann, P., Jelezko, F. & Wrachtrup, J. Photo-induced ionization dynamics of the nitrogen vacancy defect in diamond investigated by single-shot charge state detection. *New Journal of Physics* **15**, 013064 (2013).

22. Chu, Y. *et al.* Coherent Optical Transitions in Implanted Nitrogen Vacancy Centers. *Nano Letters* **14**, 1982–1986 (2014).

23. Waldherr, G. *et al.* Quantum error correction in a solid-state hybrid spin register. *Nature*

**506**, 204–207 (2014).

24. Saeedi, K. *et al.* Room-temperature quantum bit storage exceeding 39 minutes using ionized donors in silicon-28. *Science* **342**, 830–833 (2013).

25. Pfender, M. *et al.* Protecting a Diamond Quantum Memory by Charge State Control. *Nano Lett.* **17**, 5931–5937 (2017).

26. Szász, K. *et al.* Spin and photophysics of carbon-antisite vacancy defect in 4H silicon carbide: A potential quantum bit. *Phys. Rev. B* **91**, 121201(R) (2015).

27. Wolfowicz, G. *et al.* Optical charge state control of spin defects in 4H-SiC. *Nat. Commun.* **8**, 1876 (2017).

28. Dhomkar, S., Henshaw, J., Jayakumar, H. & Meriles, C. A. Long-term data storage in diamond. *Sci Adv* **2**, e1600911 (2016).

29. Broadway, D. A. *et al.* Spatial mapping of band bending in semiconductor devices using in situ quantum sensors. *Nature Electronics* **1**, 502 (2018).

30. Casas, C. F. de las *et al.* Stark tuning and electrical charge state control of single divacancies in silicon carbide. *Applied Physics Letters* **111**, 262403 (2017).

31. Christle, D. J. *et al.* Isolated Spin Qubits in SiC with a High-Fidelity Infrared Spin-to-Photon Interface. *Phys. Rev. X* **7**, 021046 (2017).

32. Widmann, M. *et al.* Coherent control of single spins in silicon carbide at room temperature. *Nat. Mater.* **14**, 164–168 (2015).

33. Castelletto, S. *et al.* A silicon carbide room-temperature single-photon source. *Nat. Mater.* **13**, 151–156 (2014).

34. Khramtsov, I. A., Vyshnevyy, A. A. & Fedyanin, D. Y. Enhancing the brightness of

electrically driven single-photon sources using color centers in silicon carbide. *npj Quantum Information* **4**, 15 (2018).

35. Simin, D. *et al.* Locking of electron spin coherence above 20 ms in natural silicon carbide. *Physical Review B* **95**, 161201 (2017).

36. Evwaraye, A. O., Smith, S. R. & Mitchel, W. C. Shallow and deep levels in n-type 4H-SiC. *J. Appl. Phys.* **79**, 7726–7730 (1996).

37. Stenger, I. *et al.* Impurity-to-band activation energy in phosphorus doped diamond. *Journal of Applied Physics* **114**, 073711 (2013).

38. Dimitrijev, S. Silicon carbide as a material for mainstream electronics. *Microelectron. Eng.* **83**, 123–125 (2006).

39. Simin, D. *et al.* High-Precision Angle-Resolved Magnetometry with Uniaxial Quantum Centers in Silicon Carbide. *Phys. Rev. Applied* **4**, 014009 (2015).

40. Niethammer, M. *et al.* Vector Magnetometry Using Silicon Vacancies in 4H-SiC Under Ambient Conditions. *Phys. Rev. Applied* **6**, 034001 (2016).

41. Anisimov, A. N. *et al.* Optical thermometry based on level anticrossing in silicon carbide. *Sci. Rep.* **6**, 33301 (2016).

42. Soltamov, V. A., Soltamova, A. A., Baranov, P. G. & Proskuryakov, I. I. Room temperature coherent spin alignment of silicon vacancies in 4H- and 6H-SiC. *Phys. Rev. Lett.* **108**, 226402 (2012).

43. Nagy, R. *et al.* Quantum Properties of Dichroic Silicon Vacancies in Silicon Carbide. *Physical Review Applied* **9**, 034022 (2018).

44. Nagy, R. *et al.* High-fidelity spin and optical control of single silicon-vacancy centres in

silicon carbide. *Nat. Commun.* **10**, 1954 (2019).

45. Banks, H. B. *et al.* Resonant Optical Spin Initialization and Readout of Single Silicon Vacancies in 4H-SiC. *Phys. Rev. Applied* **11**, 024013 (2019).

46. Magnusson, B. *et al.* Excitation properties of the divacancy in 4H-SiC. *Phys. Rev. B Condens. Matter* **98**, 195202 (2018).

47. Ivády, V. *et al.* Identification of Si-vacancy related room-temperature qubits in 4H silicon carbide. *Phys. Rev. B* **96**, 161114 (2017).

48. Trinh, X. T. *et al.* Negative-U carbon vacancy in 4H-SiC: Assessment of charge correction schemes and identification of the negative carbon vacancy at the quasicubic site. *Phys. Rev. B* **88**, 235209 (2013).

49. Fedyanin, D. Y. & Agio, M. Ultrabright single-photon source on diamond with electrical pumping at room and high temperatures. *New J. Phys.* **18**, 073012 (2016).

50. Khramtsov, I. A., Agio, M. & Fedyanin, D. Y. Dynamics of Single-Photon Emission from Electrically Pumped Color Centers. *Phys. Rev. Applied* **8**, 024031 (2017).

51. Khramtsov, I. A., Agio, M. & Yu. Fedyanin, D. Kinetics of single-photon emission from electrically pumped NV centers in diamond. *AIP Conf. Proc.* **1874**, 040014 (2017).

52. Booker, I. D. *et al.* Donor and double-donor transitions of the carbon vacancy related EH6/7 deep level in 4H-SiC. *J. Appl. Phys.* **119**, 235703 (2016).

53. Morfa, A. J. *et al.* Single-Photon Emission and Quantum Characterization of Zinc Oxide Defects. *Nano Letters* **12**, 949–954 (2012).

54. Berhane, A. M. *et al.* Single-Photon Emission: Bright Room-Temperature Single-Photon Emission from Defects in Gallium Nitride (Adv. Mater. 12/2017). *Advanced Materials* **29**,

(2017).

55. MacQuarrie, E. R., Gosavi, T. A., Jungwirth, N. R., Bhave, S. A. & Fuchs, G. D. Mechanical spin control of nitrogen-vacancy centers in diamond. *Phys. Rev. Lett.* **111**, 227602 (2013).

56. Baranov, P. G. *et al.* Silicon vacancy in SiC as a promising quantum system for single-defect and single-photon spectroscopy. *Phys. Rev. B* **83**, 125203 (2011).

57. Son, N. T., Carlsson, P., ul Hassan, J., Magnusson, B. & Janzén, E. Defects and carrier compensation in semi-insulating 4H−SiC substrates. *Phys. Rev. B* **75**, 155204 (2007).

58. Bockstedte, M., Schütz, F., Garratt, T., Ivády, V. & Gali, A. Ab initio description of highly correlated states in defects for realizing quantum bits. *npj Quantum Materials* **3**, 31 (2018).

59. Hornos, T., Gali, A. & Svensson, B. G. Large-Scale Electronic Structure Calculations of Vacancies in 4H-SiC Using the Heyd-Scuseria-Ernzerhof Screened Hybrid Density Functional. *Mater. Sci. Forum* **679-680**, 261–264 (2011).

60. Bockstedte, M., Marini, A., Pankratov, O. & Rubio, A. Many-body effects in the excitation spectrum of a defect in SiC. *Phys. Rev. Lett.* **105**, 026401 (2010).

# Supplementary Information for

"Electrical charge state manipulation of single silicon vacancies in a silicon carbide quantum optoelectronic device"

## Supplementary Note 1. Experimental Setup

All measurements were performed in a typical home-built confocal microscopy setup (see Supplementary Figure 1). Laser beams (diodes and Ti:Sapphire laser) are sent through an acoustic optical modulator (AOM) allowing for laser intensity tuning and pulsing with 10ns rise time. A dichroic mirror (DCM) transmits the excitation laser towards the objective (Olympus, UPLSAPO 60XO, oil immersion, NA 1.35), while reflecting the emitter fluorescence towards the detection system. Fluorescence is spatially filtered by a pinhole (PH) (70 µm diameter), and re-collimated. A half-wave plate (HWP) (Thorlabs AHWP10M-980) is used to equilibrate the light at the two output ports at the polarizing beam splitter (PBS) before the photon detectors. Two single photon counting avalanche photodiodes (APD) (PerkinElmer SPCM-AQRH-15) in a Hanbury-Brown and Twiss configuration (HBT) [1] were used for detection. Glan-Taylor polarizers (GTP) between the APDs and the PBS are used to block the breakdown emissions from Si APDs [2]. Through the entire work, strong phonon sideband of the V2 centre PL spectrum in the range between 900 and 1050 nm are integrated.

A coated copper wire (50 µm diameter) on top of the sample allows for applying radio-frequency signals for spin driving. The radio-frequency (RF) signal is generated by a ROHDE&SCHWARZ SMIQ03B generator and subsequently amplified by a broadband amplifier (Minicircuits ZHL-42W). Wire coating prevents shortenings between the RF wire and the top electrode. Both, top and bottom electrodes of the pin-sample are connected to a source and measurement (SMU) unit (Keithley 4200SCS).

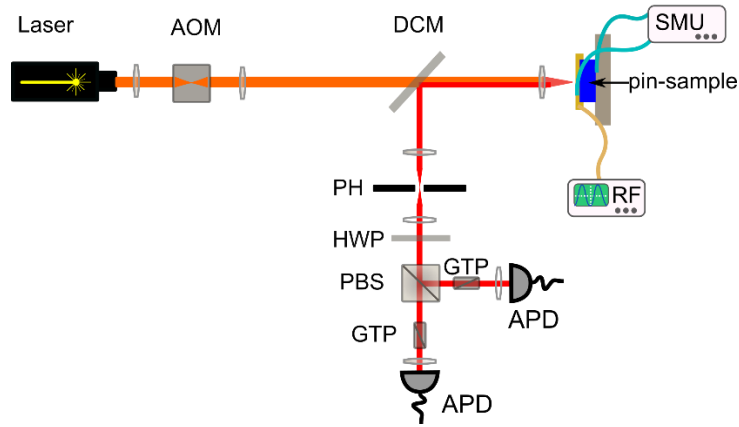

**Supplementary Figure 1**. Experimental Setup. See text for details.

## Supplementary Note 2. Identification of the emitters as $V_{Si}$

As the $V_{Si}$ in 4H-SiC does not show a characteristic zero phonon line at room temperature, their identification as $V_{Si}$ from their photoluminescence spectrum is speculative [3]. In order to prove that the observed single emitters are $V_{Si}$ we used spin resonance experiments. Coherent Rabi oscillations, obtained by the method explained elsewhere [2], with a radio-frequency driving at 70.6 MHz are reported in Supplementary Figure 2. The driving frequency corresponds to the zero-field splitting of the V2 form of $V_{Si}$ at zero magnetic field [2,4].

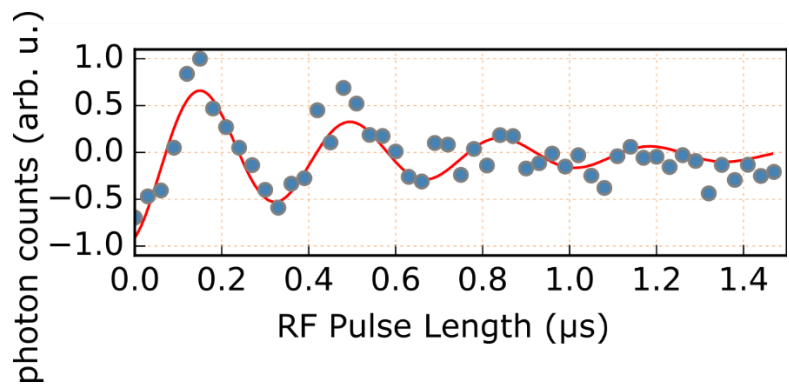

**Supplementary Figure 2.** Optically detected spin-Rabi oscillations of a single $V_{Si}$ plotted as a function the RF pulse length. This is as shown in Figure 2 (d) of the main text.

## Supplementary Note 3. $V_{Si}$ creation by electron irradiation

Silicon vacancies were created by 2 MeV electron irradiation on a 50 μm thick CVD grown 4H-SiC single crystal sample. Electron irradiation creates defects homogeneously through the whole sample volume. Water cooling during electron irradiation keeps the sample temperature close to room temperature to avoid unwanted annealing, which could convert silicon vacancies to other types of defects, such as divacancies [5]. The defect concentration increases with increasing radiation dose as shown in Supplementary Figure 3. The defect concentration is obtained as follows. First, the average total intensity from a single silicon vacancy centre was determined for a given laser power. Next, the total number of created defect centers were counted by normalizing the total fluorescence intensity of a wide two-dimensional confocal fluorescence scanning by the obtained average intensity. The result is plotted in Supplementary Figure 3, showing a linear dependence between the concentration of the silicon vacancy centres and the used electron fluence. The linear dependence can be described the empirical formula [3]:

$$\eta = \zeta e^{\alpha} \qquad (1)$$

where $\eta$ is the concentration of single emitters, $e$ the electron irradiation fluence and $\alpha = 0.47$, which was set as a free fit parameter.

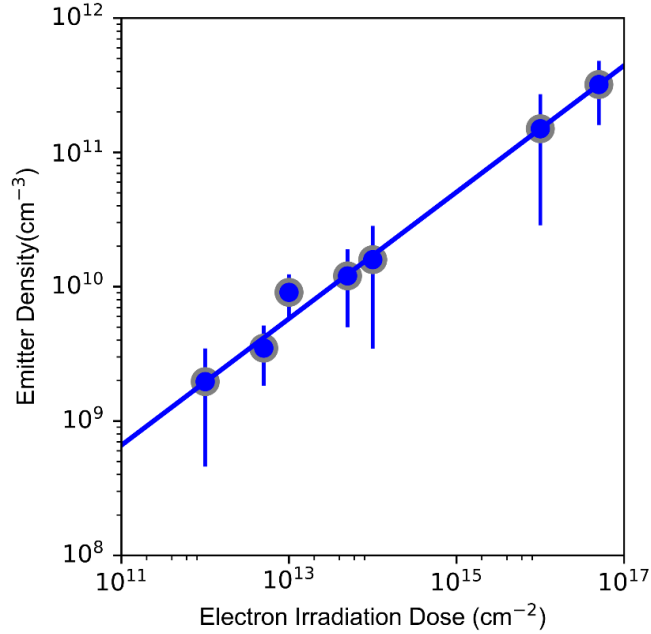

**Supplementary Figure 3.** $V_{Si}$ density plotted as a function of the irradiation dose of the 2 MeV electrons. The solid line is a fit function.

## Supplementary Note 4. Depth determination in a high refractive index material

For an emitter embedded in a high refractive-index material, such as diamond or silicon carbide, the focus of the objective is shifted deeper into the crystal and the wavefront point spread function (PSF) is distorted [6], for both the excitation laser and photoluminescence collection To precisely determine the depth of emitters observed by confocal microscopy, both excitation depth and distortion of PSF for emitters need to be considered. In the experimental setup used in this study, the emitter depth is optically determined by the excitation depth because the diameter of the excitation laser is small compared to the diameter of the back focal plane of the objective. Supplementary Figure 4a shows a simplified ray optics model of an objective/immersion oil/SiC system with Gaussian-beam excitation laser. The refractive indices of the immersion oil and SiC are $n_{oil}$ = 1.5 and $n_{SiC}$ = 2.6, respectively. In this model, we neglect the polarization of the laser.

The intensity distribution of the excitation laser is a function of radial distance as

$$I(r) = I_0 \exp\left(-\frac{r^2}{r_0^2}\right)$$

(2)

Here, $r_0$ can be calculated from full width at half maximum (FWHM) as $r_0 = (\text{FWHM})/2\sqrt{\ln 2}$. The laser light is focused through an objective with a numerical aperture NA and the diameter of the back focal plane $D$. The ray of the laser beam at the radial distance enters the SiC crystal with the incident angle $\theta_{oil}$ and refracted. The angle of refraction $\theta_{SiC}$ can be calculated from Snell's Law. Assuming the objective satisfies Abbe's sine condition, $\theta_{oil}$ satisfies

$$r = \frac{D}{2}\frac{\sin\theta_{oil}}{\sin\theta_{oil}^{max}}$$

(3)

where $\theta_{oil}^{max} = \arcsin[(\text{NA})/n_{oil}]$ is the maximum incident angle. Using Eqs. (2) and (3), the intensity distribution can be written as a function of $\theta_{oil}$: $I(\theta_{oil})$. The laser focuses at the depth of $d_{oil}$ when the SiC crystal is not placed. The differential power at the focus point contributed from the rays with the incident angle $\theta_{oil}$ is given by

$$\frac{dP}{d\theta_{oil}} = I(\theta_{oil}) \cdot \frac{d}{d\theta_{oil}}(\pi r^2) \cdot T(\theta_{oil})$$

(4)

Here, $T(\theta_{oil})$ is the Fresnel transmission coefficient for non-polarized light

$$T(\theta_{\text{oil}}) = \frac{2n_{\text{SiC}}n_{\text{oil}}\cos\theta_{\text{SiC}}\cos\theta_{\text{oil}}}{(n_{\text{SiC}}\cos\theta_{\text{oil}} + n_{\text{oil}}\cos\theta_{\text{SiC}})^2} + \frac{2n_{\text{SiC}}n_{\text{oil}}\cos\theta_{\text{SiC}}\cos\theta_{\text{oil}}}{(n_{\text{SiC}}\cos\theta_{\text{SiC}} + n_{\text{oil}}\cos\theta_{\text{oil}})^2}$$

(5)

Because the depth of the focus inside SiC is given by [6]

$$d_{\text{SiC}}(\theta_{\text{oil}}) = \frac{n_{\text{SiC}}\cos\theta_{\text{SiC}}}{n_{\text{oil}}\cos\theta_{\text{oil}}}d_{\text{oil}}$$

(6)

the laser power density as a function of $d_{\text{SiC}}$ can be obtained from Eq. (5) and (6) using a parameter $\theta_{\text{oil}}$. The average of $d_{\text{SiC}}$ by the power density Eq. (4) gives the depth correction factor limited by excitation:

$$\langle d_{\text{SiC}} \rangle = \int_0^{\theta_{\text{oil}}^{\max}} d_{\text{SiC}} \frac{\mathrm{d}P}{\mathrm{d}\theta_{\text{oil}}}\mathrm{d}\theta_{\text{oil}} \bigg/ \int_0^{\theta_{\text{oil}}^{\max}} \frac{\mathrm{d}P}{\mathrm{d}\theta_{\text{oil}}}\mathrm{d}\theta_{\text{oil}}$$

(7)

The back aperture diameter of the used objective is 8.1 mm, and the FWHM of the laser is approximately 2–2.5 mm. Under this condition, $\langle d_{SiC}\rangle \simeq 1.9 d_{oil}$ is obtained. This value almost equals to the lower limit of $d_{SiC}/d_{oil}$ determined by parallel rays: $\min(d_{SiC}/d_{oil}) = n_{SiC}/n_{oil}$. Supplementary Figure 4b shows the calculation result of $\mathrm{d}P/\mathrm{d}(d_{SiC}/d_{oil})$ as a function of $d_{SiC}/d_{oil}$ at various FWHM of the excitation laser beam. In the experimental condition in this study, it can be seen that most of the excitation laser is focused at the depth of $1.8d_{\text{oil}}$. Therefore it is expected that the emitters at the depth of $1.8d_{\text{oil}}$ are observed.

The correction factor is used to correct for all measured emitter depths inside the pin-diode. One example is shown in Supplementary Figure 5 obtained from another device in which high density $V_{Si}$ ensembles are created by high dose electron irradiation ($1e16/cm^2$). In Supplementary Figure 5 b,c, the bright layer at $z = 0$ is likely related to the indium tin oxide (ITO) layer, as this layer is not visible on a different sample without ITO. A bright layer of the $V_{Si}$ ensemble is found by moving the sample stage towards the SiC sample roughly 26 µm which is equivalent to 47 µm in SiC according to the correction factor obtained above. This is close to the depth of the i-n interface estimated by the growth condition.

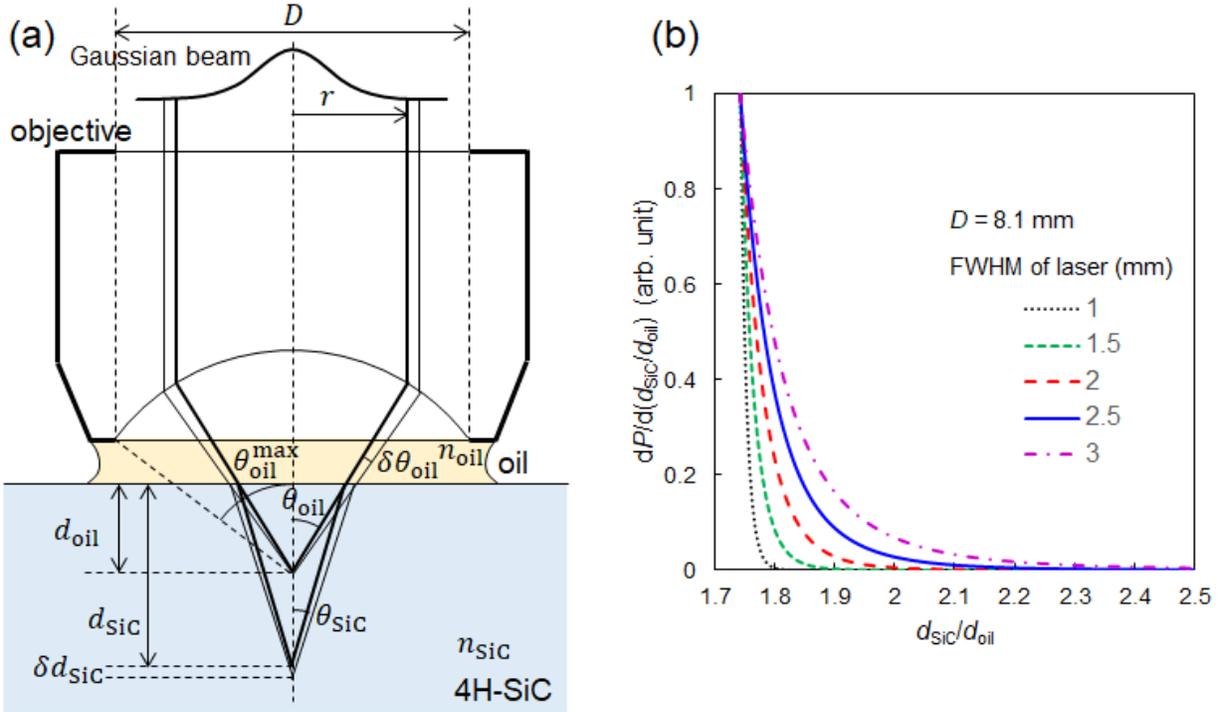

**Supplementary Figure 4.** (a) Schematic of the ray optics model for laser excitation at objective/immersion oil/SiC system. (b) Calculated depth dependence of laser power distribution inside SiC crystal at various FWHM of excitation laser beam.

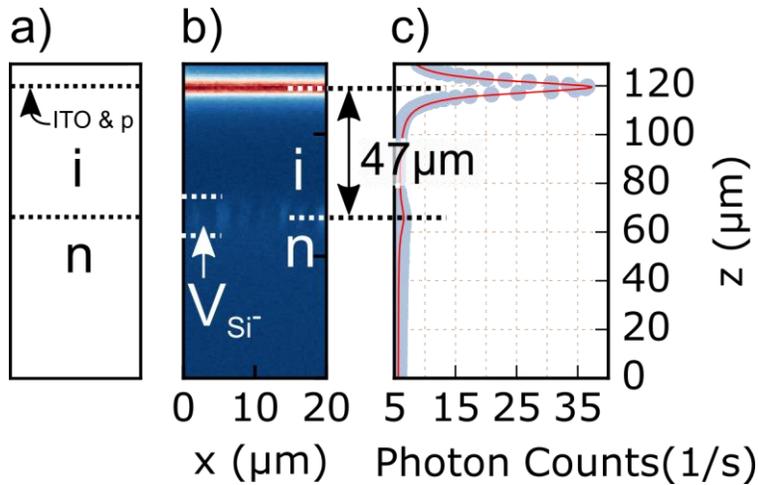

**Supplementary Figure 5.** (a) Schematic of the used pin diode. (b) Confocal raster scan through vertical cross section of the device irradiated with electrons with a dose of 5E12 cm$^{-2}$. (c) Integrated PL intensity taken from panel (b).

# Supplementary Note 5. Additional data for charge state switching

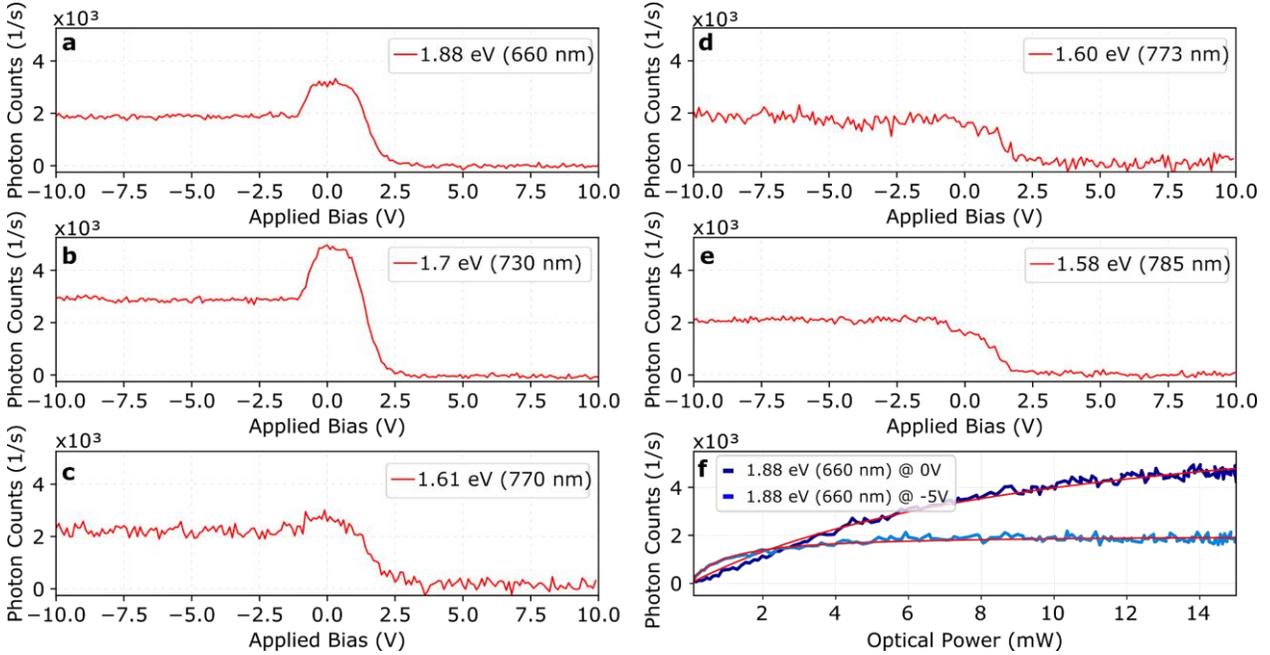

**Supplementary Figure 6. Optical excitation dependence of charge state conversion.** (a-e) Bias dependent PL intensity curves under different excitation energies $\hbar\omega$: 1.88 eV (660 nm) (a), 1.70 eV (730 nm) (b), 1.62 eV (765.5 nm) (c), 1.60 eV (773 nm) (d), and 1.58 eV (785 nm) (e). (f) Optical saturation curves measured at $\hbar\omega$=1.88 eV and $V$ = -5 V (bright blue curve) and $V$ = 0 V (dark blue curve). Fit function: $I = \alpha P/(\beta + P)$, where $\alpha$ specifies the saturated count rate and $\beta$ the saturation power are fit parameters. At 0 V $\alpha$=5.2±0.3 kcts and $\beta$=10±1 mW, and at -5V $\alpha$=2.0±0.3 kcts and $\beta$=0.90±0.01 mW. This is as shown in the main figure 3d.

To characterize how optical excitation is related with the observations in Fig. 1 and 2 of the main text, we test many $V_{Si}$ defects and monitor their PL intensity as a function of the applied voltage while sweeping the optical excitation energy $\hbar\omega$. In Supplementary Figure 6a, the integrated PL intensity for the selected single $V_{Si}$ near the i-n interface is plotted versus the bias voltage for $\hbar\omega$ = 1.88 eV ($\lambda$ = 660 nm). We vary the wavelength of the pump laser. Supplementary Figure 6b,c show that similar curves are obtained at $\hbar\omega$ > 1.60 eV ($\lambda$ < 773 nm). However, this peak in the PL intensity at the switching bias disappears at $\hbar\omega \lesssim$ 1.60 eV ($\lambda \gtrsim$ 773 nm) (Supplementary Figure 6d,e). In the main text figure 3, data in b, c, and d are shown. We also measured optical saturation curves at the switching bias voltage (0 V) and a lower voltage (-5 V), which are shown in Supplementary Figure 6f. It appears that a stronger optical excitation power is necessary at the switching bias to optically saturate the $V_{Si}^{(-)}$.

# Supplementary Note 6. Calculated optical absorption spectra and optical ionization of the silicon vacancy

As outlined in the main text the photo ionization has impact on the PL intensity of the silicon vacancy. In this section we address the optical absorption spectra of the silicon vacancy as calculated for the negative charge state $V_{Si}^{(-)}$ as well as for the neutral and doubly negative charge states that result from the ionization of $V_{Si}^{(-)}$. The vertical absorption spectra as shown in Supplementary Figure 7 were obtained using the CI-CRPA method as outlined in the theoretical method section and refer to excitations of the defect in the ground state geometry of the corresponding charge states.

**Theoretical Methods.** The electronic and optical properties of the silicon vacancy in 4H-SiC were investigated in the framework of density functional theory and the configuration interaction method CI-CRPA [7]. The latter is employed to describe the highly correlated multiplet states of the silicon vacancy and to calculate optical absorption spectra for excitations among the multiplet states or between multiplet states and extend band states. The CI-CRPA method uses the geometry obtained within DFT and the hybrid functionals HSE06 [8] as implemented in the VASP program package. The CI basis set is constructed from Kohn-Sham orbitals including as a minimal set the defect states in the band gap, a few valence band states including defect resonances, and for the calculation of optical absorption spectra additional valence and conduction band states. In the construction of the CI-basis set we include all possible excitation between the localized defect states and major defect resonances in a range from the valence band edge $E_V$ down to $E_V$ - 1 eV. Conduction bands from the conduction band edge up to $E_C$ + 1.2 eV were included for calculation of optical absorption spectra. Additional states were included via single excitation. First all single excitations are considered starting from the high spin ground state and subsequently slater determinants corresponding to higher excitation are generated from these states. The silicon vacancy in 4H-SiC is represented by a large super cell with 576 lattice sites using an energy cut-off of 400 eV for augmented plane wave basis set. In 4H-SiC the silicon vacancy can be created at the two inequivalent lattice sites, namely the hexagonal and cubic sites. Both defects possess very similar optical spectra with slightly different PL peaks. For simplicity we show spectra for one silicon vacancy. Optical absorption spectra are calculated as vertical excitations from the ground state geometry of the defect as described in Ref. [7].

**Results.** The multiplet states of the hexagonal $V_{Si}^{(-)}$ in its ground state geometry as obtained with the CI-CRPA approach is shown in Supplementary Figure a. The lowest fundamental quartet excited state is formed $^4A'_2$ and $^4E$ quartet (at 1.5 and 1.57 eV respectively). In between the ground state and excited state quartet we find a group of nearly degenerate doublets $^2A_2$, $^2E$, an $^2E'$ doublet and a group of nearly degenerate doublets $^2A_1$, $2E''$. The latter is found just below the fundamental excited quartet and mediates the spin-selective transitions via inter system crossings.

The energetic ordering of quartet and doublet levels derived here from a first principles approach verifies the ordering suggested based on a group theoretical treatment and qualitative arguments [9]. The calculated zero phonon lines (ZPLs) for the transition from the lowest excited quartet states to the ground state of the hexagonal $V_{Si}^{(-)}$ are found at 1.35 and 1.39 eV. For the cubic $V_{Si}^{(-)}$ the same multiplet ordering and the ZPLs at 1.25 and 1.32 eV are obtained. The splitting between $^4A'_2$ and $^4E$ ZPLs is determined by a (pseudo) Jahn-Teller effect and is overestimated due to the finite size of our already large supercells. Recently, the hexagonal (cubic) vacancy was identified with the V1/V1' (V2) centers [10]. In experiment [10] the Si-vacancy related ZPLs V1 and V1' are found at 1.439 eV and 1.445 eV and V2 at 1.35 eV. The calculated values for the two defects are in agreement with experiment given the consistent underestimation of the ZPLs by our CI-CRPA approach of about 100 meV that was also observed for other defects [7]. Calculated absorption spectra (cf. Supplementary Figure 7b for the vertical adsorption of the hexagonal vacancy) show the same polarization for the allowed transition to $^4A_2$ ($^4E$) as found for the V1 (V1') ZPL or the V2 ZPL. These polarization properties are also obtained for the distorted defect in the $^4E$ quartet excited geometries – given that the defect in this geometry possesses only $C_{1h}$ symmetry, our result is not granted by symmetry.

We now turn to the photo-ionization. The vertical absorption spectrum of the hexagonal $V_{Si}^{(-)}$ in Supplementary Figure 7b also indicates the excitation threshold for the onset of excitations from the valence band to defect levels or from the defect levels to conduction band states. The former transitions lead to the ionization $V_{Si}^{(-)} \rightarrow V_{Si}^{(2-)} + h^+$, which produces a hole in the valence band and has an excitation threshold of 1.96 eV (1.98 eV cubic $V_{Si}^{(-)}$). The latter transitions produce the neutral vacancy and an electron in the conduction band, $V_{Si}^{(-)} \rightarrow V_{Si}^{(0)} + e^-$, with a threshold of 2.1 eV (2.2 eV for cubic $V_{Si}^{(-)}$). As indicated in Supplementary Figure 7b the dominant absorption above 2 eV leads to an ionization to $V_{Si}^{(2-)}$ while the formation of $V_{Si}^{(0)}$ plays a minor role. Nevertheless, the photon energy employed in the experiments for the excitation of the $V_{Si}^{(-)}$ PL is not sufficient to ionize $V_{Si}^{(-)}$, instead, non-radiative recombination with electrons or holes are the relevant processes that establishes a steady state.

In contrast to $V_{Si}^{(-)}$, photo ionization of $V_{Si}^{(0)}$ and $V_{Si}^{(2-)}$ is enabled at photon energies employed in our experiments. The ionization thresholds that yield $V_{Si}^{(-)}$ are indicated in the vertical adsorption spectra shown in Supplementary Figure 7c and d. The neutral vacancy $V_{Si}^{(0)}$ possess a triplet ground state $^3E$ that is almost degenerate with a singlet $^1E$ for both kind of vacancies. The vertical optical ionization threshold for the ionization $V_{Si}^{(0)} \to V_{Si}^{(-)}$ is found at 0.9 eV. At larger excitation energies between 1.6 and 1.9 eV a strong peak arises from a coupling of valence band states and defect resonances in the excitation of both kind of vacancies in the triplet ground state. Such a peak is not observed for the excitations of the singlet state as indicated in Supplementary Figure . For the doubly negatively charged vacancy $V_{Si}^{(2-)}$ a triplet ground state $^3A_2$ is obtained. The vertical absorption spectrum shows intra defect excitations at 0.4 and 1.6 eV. The vertical optical ionization threshold is found at 1.0 eV. Thus under the experimental conditions optical conversion of $V_{Si}^{(0)}$ and $V_{Si}^{(2-)}$ to $V_{Si}^{(-)}$ is an active process, that reinitializes $V_{Si}^{(-)}$.

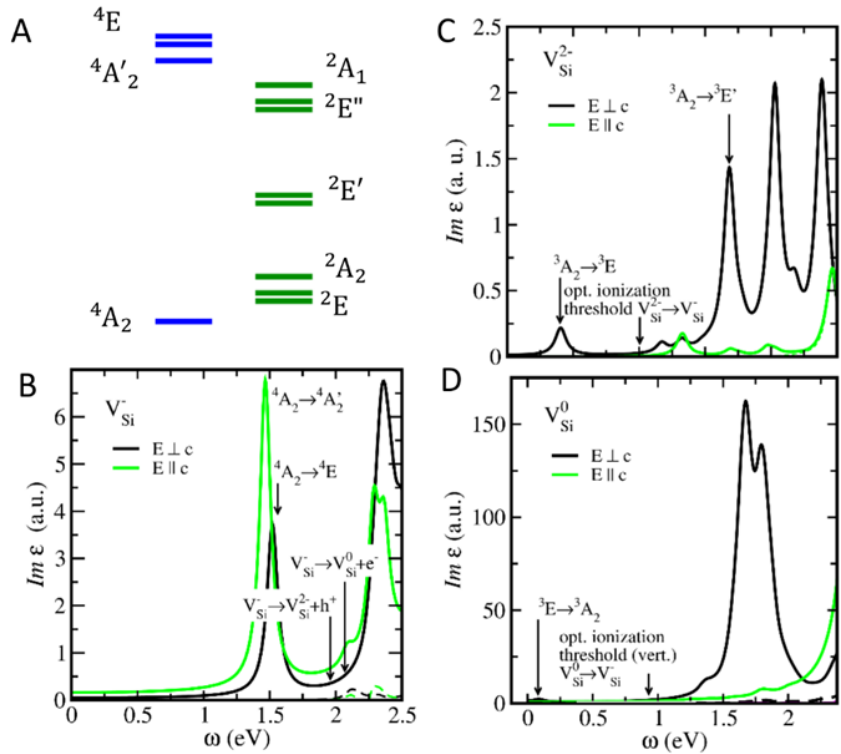

**Supplementary Figure 7.** Optical transitions of the silicon vacancy: (a) multiplet states $V_{Si}^{(-)}$. Optical absorption spectra of (b) the negative vacancy $V_{Si}^{(-)}$, (c) of $V_{Si}^{(-2)}$, and (d) of $V_{Si}^{(0)}$ as obtained with the CI-CRPA at the hexagonal site – spectra at the cubic site are similar and only differ by the slightly different excitation/threshold energies as indicated in the text. Individual transition are broadened by 50 meV. Internal optical transitions and the thresholds for optical ionization are indicated by arrows. For $V_{Si}^{(-)}$ excitation of electrons from the valence band (i.e. $V_{Si}^{(-)} \rightarrow V_{Si}^{(2-)}$) dominates the ionization via the conduction band (dashed lines in (b)). Note, that the ionization threshold of $V_{Si}^{(0)}$ in (d) occurs at much lower energies than the main peak related to the excitation of the triplet ground state – dashed lines correspond to excitations from the lowest singlet state.

## Supplementary Note 7. IV characteristics of the p-i-n diode

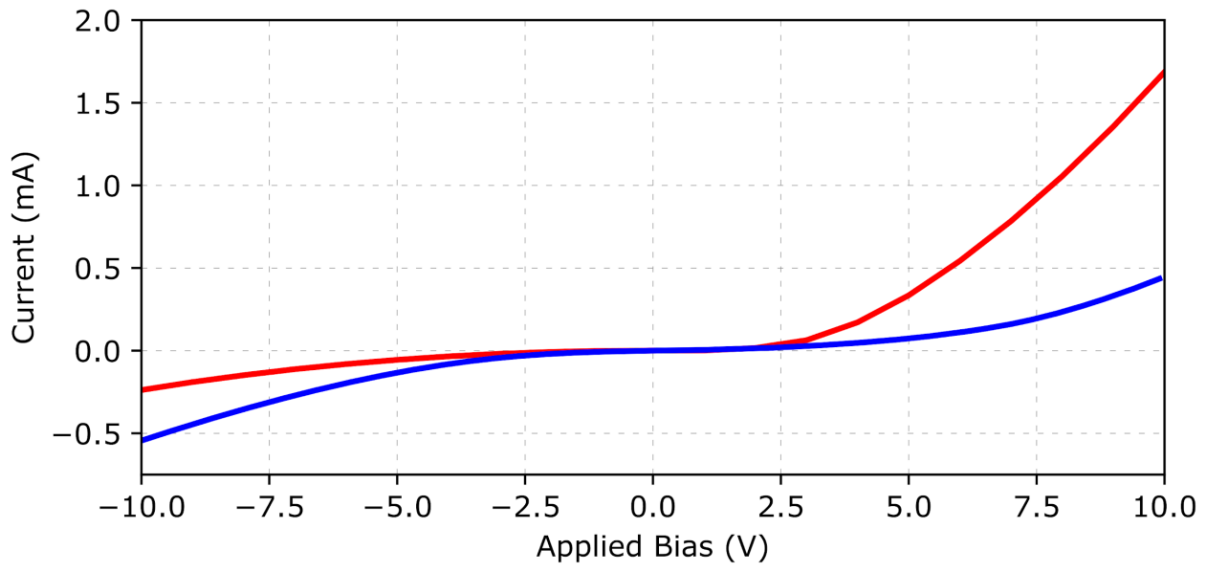

**Supplementary Figure 8.** IV curves of the used p-i-n diode device measured in 2015 (red) and in 2017 (blue). The difference in the measured IV curves is most likely due to the device degradation. Since most of the data remeasured in 2017 are shown in this manuscript, the resistance at high forward and reverse bias, 2 and 8 kΩ, respectively, are used for device simulation as explained in Methods of the main text.

# Supplementary Note 8. Numerical simulation of the p-i-n diode

**Supplementary Table 1.** Parameters used in the numerical simulations.

| Parameter | Value | Comment |
|---|---|---|
| Energy band gap of 4H-SiC at 300 K | 3.26 eV | Ref. [11] |
| Energy band gap of ITO at 300 K | 3.9 | Ref. [12] |
| Dielectric constant of 4H-SiC | 9.8 | Ref. [13] |
| Dielectric constant of ITO | 9.0 | |
| Activation energy of donors at cubic sites | 0.10 eV | Ref. [14] |
| Activation energy of donors at hexagonal sites | 0.05 eV | Ref. [14] |
| Activation energy of acceptors | 0.19 eV | Ref. [15] |
| DOS effective mass for electrons | $0.77 m_0$ | Ref. [16] |
| DOS effective mass for holes | $0.91 m_0$ | Ref. [17] |
| Electron density in the conduction band of the ITO layer | $8.5 \times 10^{20}$ cm$^{-3}$ | Ref. [12] |
| Acceptor concentration in the p$^{++}$-type layer | $1 \times 10^{19}$ cm$^{-3}$ | |
| Acceptor compensation ratio in the p$^{++}$-type layer | 5% | |
| Acceptor concentration in the p-type layer | $1 \times 10^{18}$ cm$^{-3}$ | |
| Acceptor compensation ratio in the p-type layer | 5% | |
| Acceptor concentration in the i-region | $3.2 \times 10^{13}$ cm$^{-3}$ | |
| Donor concentration in the i-region | $3.2 \times 10^{12}$ cm$^{-3}$ | |
| Donor concentration in the n-type layer | $1 \times 10^{18}$ cm$^{-3}$ | |
| Donor compensation ratio in the n-type layer | 5% | |
| Donor concentration in the n$^{++}$-substrate | $9 \times 10^{18}$ cm$^{-3}$ | |
| Donor compensation ratio in the n$^{++}$-substrate | 5% | |
| Electron mobility in the ITO layer | 12.8 cm$^2$/Vs | Ref. [12] |
| Hole mobility in the ITO layer | 12.8 cm$^2$/Vs | Ref. [12] |
| Electron mobility in the p$^+$-type region | 170 cm$^2$/Vs | Refs. [18,19] |
| Hole mobility in the p$^+$-type region | 90 cm$^2$/Vs | Refs. [18,19] |
| Electron mobility in the p-type region | 440 cm$^2$/Vs | Refs. [18,19] |
| Hole mobility in the p-type region | 110 cm$^2$/Vs | Refs. [18,19] |
| Electron mobility in the i-type region | 900 cm$^2$/Vs | Refs. [18,19] |
| Hole mobility in the i-type region | 115 cm$^2$/Vs | Refs. [18,19] |
| Electron mobility in the n-type region | 280 cm$^2$/Vs | Refs. [18,19] |
| Hole mobility in the n-type region | 100 cm$^2$/Vs | Refs. [18,19] |
| Electron mobility in the p$^+$-type region | 100 cm$^2$/Vs | Refs. [18,19] |
| Hole mobility in the p$^+$-type region | 80 cm$^2$/Vs | Refs. [18,19] |

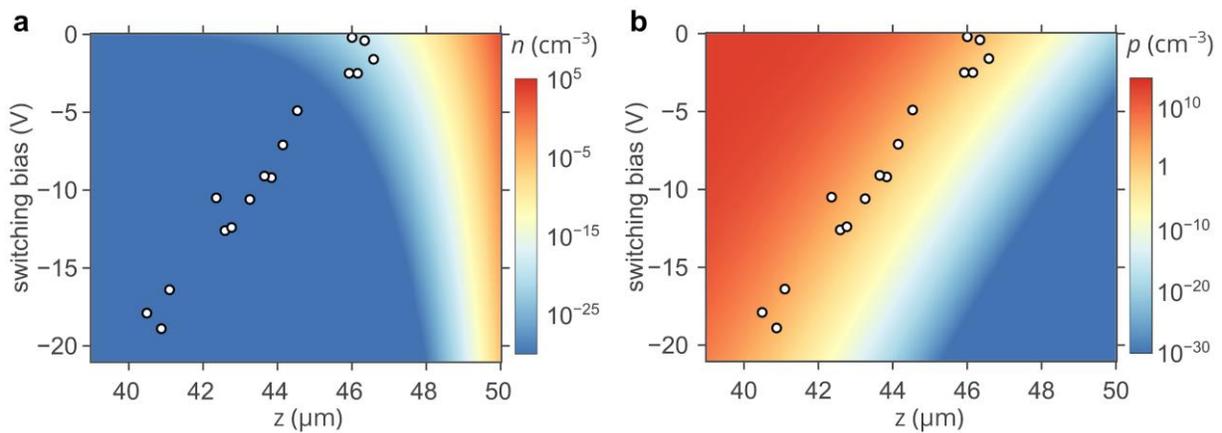

**Supplementary Figure 9.** Evolution of the spatial distributions of the electron (panel a) and hole (panel b) densities with the reverse voltage applied to the device. Open dots represent the results of the experimental measurements shown in Fig. 2c in the main text.

# References

1. Brown, R. H. & Twiss, R. Q. LXXIV. A new type of interferometer for use in radio astronomy. *The London, Edinburgh, and Dublin Philosophical Magazine and Journal of Science* **45**, 663–682 (1954).

2. Widmann, M. *et al.* Coherent control of single spins in silicon carbide at room temperature. *Nat. Mater.* **14**, 164–168 (2015).

3. Fuchs, F. *et al.* Engineering near-infrared single-photon emitters with optically active spins in ultrapure silicon carbide. *Nat. Commun.* **6**, 7578 (2015).

4. Niethammer, M. *et al.* Vector Magnetometry Using Silicon Vacancies in 4H-SiC Under Ambient Conditions. *Phys. Rev. Applied* **6**, 034001 (2016).

5. Christle, D. J. *et al.* Isolated electron spins in silicon carbide with millisecond coherence times. *Nat. Mater.* **14**, 160–163 (2015).

6. MacQuarrie, E. R., Gosavi, T. A., Jungwirth, N. R., Bhave, S. A. & Fuchs, G. D. Mechanical spin control of nitrogen-vacancy centers in diamond. *Phys. Rev. Lett.* **111**, 227602 (2013).

7. Bockstedte, M., Schütz, F., Garratt, T., Ivády, V. & Gali, A. Ab initio description of highly correlated states in defects for realizing quantum bits. *npj Quantum Materials* **3**, 31 (2018).

8. Krukau, A. V., Vydrov, O. A., Izmaylov, A. F. & Scuseria, G. E. Influence of the exchange screening parameter on the performance of screened hybrid functionals. *J. Chem. Phys.* **125**, 224106 (2006).

9. Soykal, Ö. O., Dev, P. & Economou, S. E. Silicon vacancy center in 4 H -SiC: Electronic structure and spin-photon interfaces. *Phys. Rev. B Condens. Matter* **93**, (2016).

10. Janzén, E. *et al.* The silicon vacancy in SiC. *Physica B* **404**, 4354–4358 (2009).

11. Abderrazak, H. & Emna Selmane Bel. Silicon Carbide: Synthesis and Properties. in *Properties and Applications of Silicon Carbide* (2011).

12. Kim, H. *et al.* Indium tin oxide thin films for organic light-emitting devices. *Appl. Phys. Lett.* **74**, 3444–3446 (1999).

13. Persson, C., Lindefelt, U. & Sernelius, B. E. Doping-induced effects on the band structure in n-type 3C−, 2H−, 4H−, 6H−SiC, and Si. *Phys. Rev. B* **60**, 16479–16493 (1999).

14. Evwaraye, A. O., Smith, S. R. & Mitchel, W. C. Shallow and deep levels in n-type 4H-SiC. *J. Appl. Phys.* **79**, 7726–7730 (1996).

15. Negoro, Y., Kimoto, T., Matsunami, H., Schmid, F. & Pensl, G. Electrical activation of high-concentration aluminum implanted in 4H-SiC. *J. Appl. Phys.* **96**, 4916–4922 (2004).

16. Son, N. T. *et al.* Electron effective masses in 4H SiC. *Appl. Phys. Lett.* **66**, 1074–1076 (1995).

17. Son, N. T. *et al.* Hole Effective Masses in 4H SiC Determined by Optically Detected Cyclotron Resonance. *Mater. Sci. Forum* **338-342**, 563–566 (2000).

18. Pernot, J. *et al.* Electrical transport in n-type 4H silicon carbide. *J. Appl. Phys.* **90**, 1869–1878 (2001).

19. Mnatsakanov, T. T., Levinshtein, M. E., Pomortseva, L. I. & Yurkov, S. N. Carrier mobility model for simulation of SiC-based electronic devices. *Semicond. Sci. Technol.* **17**, 974–977 (2002).